# Teaching Einsteinian Physics at Schools: Part 2, Models and Analogies for Quantum Physics

**Tejinder Kaur[1], David Blair[1], John Moschilla[1], and Marjan Zadnik[1]**

[1]The University of Western Australia, 35 Stirling Highway, Crawley, WA 6009, Australia.

E-mail: tkaur868@gmail.com

**Abstract**

The Einstein-First project approaches the teaching of Einsteinian physics through the use of physical models and analogies. This paper presents an approach to the teaching of quantum physics which begins by emphasising the particle-nature of light through the use of toy projectiles to represent photons. This allows key concepts including the spacing between photons, and photon momentum to be introduced. This in-turn allows an intuitive understanding of the uncertainty principle. We present optical interference in the context of individual photons, using actual videos showing the development of images one at a time. This enables simple laser interference experiments to be interpreted through the statistical arrival of photons. The wave aspects of quantum phenomenon are interpreted in terms of the wavelike nature of the arrival probabilities.

## 1. Introduction

The age of quantum physics began with two novel and revolutionary explanations for long-standing unexplained phenomena. The first was Planck's explanation of blackbody radiation in 1901, and second was Einstein's explanation of the photoelectric effect in 1905.

Heinrich Hertz's discovery of the photoelectric effect in 1887[1] puzzled the scientific community of his day. When Hertz shone light onto a metal surface, he was astounded to find that for some metals, sparks were only emitted when irradiated by light in the ultraviolet band. Light of higher frequency, rather than higher intensity, was needed to eject electrons from some metals (see activity 2.5 below). Classically the energy of light was expected to be proportional to intensity. Hertz's observations contradicted this. Eighteen years later Einstein explained the photoelectric effect by quantisation of electromagnetic radiation, consistent with Planck's earlier explanation of the blackbody radiation spectrum.[2]

Planck's explanation of the universal spectrum of blackbody radiation (emitted by hot objects) had likewise required quantisation to explain why the spectrum rolls of above a characteristic light frequency ( dependent only on the temperature). Planck had hypothesised that the radiation is emitted in discrete packets with energy proportional to frequency: $E = hf$, where h is Planck's constant and f is the frequency of light.

Richard Feynman emphasised the importance of these now century-old discoveries in his book QED: "*I want to emphasise that light comes in this form – particles. It is very important to know that light behaves like particles, especially for those of you who have gone to school, where you were probably told something about light behaving like wave*s."[3]

It is difficult to understand why the particle nature of light should not be a central concept in school education. The idea of a stream of photons that have momentum is not difficult. The concept of radiation pressure, which follows directly, has been used in the solar sail spacecraft Ikaros[4], and is seen in the tails of comets. Early introduction of the particle nature of light allows imaging such as photography to be understood in terms of the statistical arrival of individual photons and the uncertainty principle to be understood in relation to the momentum kicks provided by individual photons as discuss in Sections 2.3 and 2.4. We suggest that these concepts allow a key aspect of the quantum nature of light to be taught at the lowest levels of school science.

Many researchers are actively investigating more effective ways of teach quantum physics.[5],[6],[7] Different approaches such as computer games[8], simulations[9], multimedia[10] and virtual experiments[11] have been tested. Most research focuses on senior high school and tertiary levels.

In this paper we present an observation-based approach to conceptual understanding of quantum physics based on the use of models, analogies and thought experiments. The approach is designed to allow quantum physics to be taught in early years of school science. It involves activity based learning and minimal mathematics. It is designed to provide a foundation in quantum physics suitable for all students. In particular it is designed to answer the question "what is light" in the context of modern physics.

Today single photon sensing is commonplace, and it has recently been shown that human eyes can sense single photons.[12] Interference experiments that record individual photon arrivals show that interference patterns are formed by statistical arrival of individual photons. After enough photons have arrived the pattern looks like the pattern one might expect from interference of classical waves, despite it having been created by random arrival of single photons. The creation of an interference pattern that resembles those created by water waves tells us that the probability of photon arrival at each location is determined by the equation of a wave.

In the 19$^{th}$ century single photon observations were impossible, and observations of interference patterns conclusively supported the idea that light is a wave. However today when we view a video of interference one photon at a time, the wave-nature of light appears to be an illusion that only emerges when a large number of photons make quasi-continuous patterns that resemble the patterns created by classical waves.

In our observation-based approach, light is always a particle, but observations of interference tell us that equations of classical-type waves determine the probability that a photon arrives at a particular location. In our approach photon themselves are not some weird combination of a wave and a particle. The wave-nature of light, revealed by interference is a useful model when light intensity is high. Light is not a wave that sometimes shows particle-like behaviour but rather a stream of particles that display wave-like behaviour under certain conditions.

Our approach does not elliminate quantum weirdness, but makes it more transparent by using the concept of spacing between photons to show that photons themselves do not interfere with each other as discussed in Section 2.2. Our approach can also be easily extended to allow finite rest mass particles to be integrated into the same world view. Like photons, massive particles also share wavelength and frequency, and likewise (as observed by electrons, neutrons, atoms and buckyballs) equations of waves determine the probability of arrival in interference experiments. The fact that interference occurs when the particles arrive one at a time tells us that the interference is the interference of possible paths and not the interference of actual objects. The intrinsic mystery of why interference occurs is presented as an observational fact that we have to accept.

The conceptual approach presented here is designed to make the modern concept of light available to all, and for it to be extendable to all aspects of quantum physics. Learning about the nature of light is important not only because it is the basis of observed reality, but also because of its immense importance in modern technology. The activities presented here allow students to recognise quantum phenomena in everyday experiences. It allows their early learning to be consistent with the physics that underpins the design of the computer chips that power our laptops, desktops, tablets, smartphones, household appliances and kids toys, as well as the lasers that power our telecommunications and scanners we use to pay groceries bills. In every smartphone the navigation system relies on both quantum physics and general relativity.

This paper is designed to counter the view that quantum physics is so abstract and challenging that it is only suitable for students entering university. Indeed, when students start their university education they have difficulties because the new concepts contradict their prior knowledge of classical physics. It is inefficient and uneconomical to follow the traditional approach. Moreover it is clear that students are aware that their conventional curriculum does not relate to modern technology.

Part 1 of this series presented models and analogies to teach the concepts of relativity. The reason for using models and analogies in general relativity is that we cannot see space and visualisation of 4-dimensional space-time is very difficult. The use of models and analogies allow us to make abstract and challenging concepts tangible. Similarly, we use models and analogies to teach the concepts of quantum physics because photons are so small and quantum behaviour is invisible. To understand the concept of photons we scale up to easily visualisable objects. But these are combined with thought experiments and actual experiments with lasers.

In this paper we present a program of 6 - 10 lessons that reverses the traditional order, putting particle-like behaviour first. For comparing classical waves with patterns created by photons, we use videos of water waves and Google Earth images of waves that display diffraction and interference patterns. These compliment simple, beautiful laser interference experiments. Because single photon detection is too expensive and difficult for the classroom we use videos of single photon interference.[13] More advanced students can use the spacing-between-photons concept to realise that easily observed phenomena like interference in soap films is actually single photon interference.

Finally in section 2.9 we discuss the limitations of the models used. This is an essential part of the educational process, and can be extended into a broader debate about the nature of physical reality.

## 2. Models and analogies for teaching quantum physics

The following paragraphs describe various models and analogies to teach particle and wave nature of light.

*2.1 Analogue photons: Nerf gun projectiles*
To represent photons physically, we use small foam projectiles fired from toy guns called Nerf guns. These projectiles come with either blunt foam caps or with suction caps that can stick to surfaces such as whiteboards. We use a stream of Nerf bullets to represent a beam of light. With this analogy, we allow students to explore the following phenomena in this quantum context, 1) photography, 2) light scattering, 3) the uncertainty principle, and 4) the photoelectric effect as described in the sections below. Here we can introduce the recent discovery that humans can sense the arrival of single photons.

It is useful to preceede these activities with an investigation of the spacing between photons, as this emphasises the weirdness of quantum behaviour which underly these phenomena.

*2.2 The spacing between real photons: A thought experiment*
In this activity we use observations of projectile motion to investigate the distances between real photons. Using Nerf gun bullets and simple math, students can calculate the spacing between successive photons, with the familiar formula, $distance = speed \times time$. For example, consider a Nerf gun that fires 10 bullets per second, at a speed of 10 metres per second. Using the formula above, students can easily calculate that successive bullets are separated by a distance of 1 metre. We ask students to video a stream of bullets from a Nerf machine gun and to test the validity of that analysis.

Having introduced students to observations that a) light comes as photons and b) that the spacing between moving objects can be easily calculated when their speed and count rate (number of objects moving past a given point per second) are known, we can investigate the spacing between real photons by extending the above activity into a thought experiment.

For this thought experiment, we start with the traditional scale for the brightness of stars: zeroth magnitude is the brightest (eg. the star Sirius), down to 6$^{th}$ magnitude (just visible with the naked eye). A 6$^{th}$ magnitude star is approximately 250 times dimmer than a zeroth magnitude star. For the zeroth maginitude star, the light flux is roughly 1 million photons per second per square centimeter. We are interested to know how many photons per second enter the human eye.

The aperture of the human eye, the pupil, has a rough width of 3mm. Thus for the 6$^{th}$ magnitude star, approximately 300 photons per second enter the human eye. Applying the same logic as before (and knowing the speed of light), students can calculate the distance between successive photons:

$$Distance\ between\ successive\ photons = \frac{300\,000\,000}{300} = 1\ million\ meters$$

Thus when a photon from a 6$^{th}$ magnitude star enters our eye, the next one is 1000 km away! This remarkable result is unavoidable when we admit the particle nature of light. Even for the brightest, zeroth magnitude star, the photons are 4 km apart.

*2.3 Photography with analogue photons*
Photography using Nerf gun photons (with suction caps) is used to emphasise the particle nature of light. Our "photographs" are silhouette images created by Nerf gun bullets that stick to a white board, but not to clothing. In a real camera, photons are registered by a CCD device. In our activity, the "photons" are registered by the whiteboard (see Figure 1).

We ask students to "photograph" each other using Nerf gun bullets. One student stands against a glossy wall or whiteboard, while others "illuminate" the subject with bullets. Bullets stick only to the glossy surface, and create a silhouette "photograph" of the student as shown in Figure 1. We explain to students that photons have properties analogous to those of bullets, including energy and momentum.

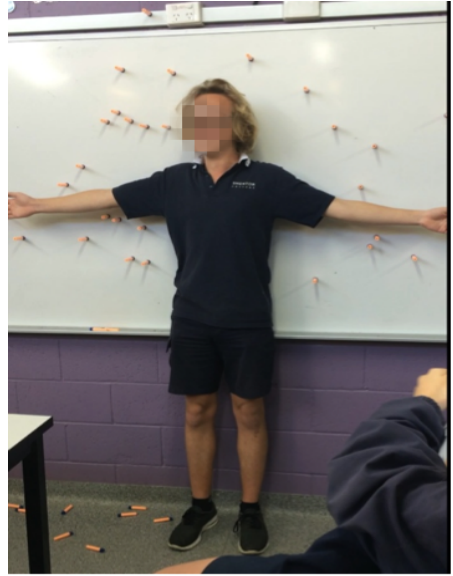

Figure 1. A 'silhouette photograph' of a student created by Nerf gun bullets, representing the linear trajectory of photons in flat space. It also demonstrates the particle-like properties of photons, namely their non-zero momentum and energy.

*2.4 Heisenberg's uncertainty principle with analogue photons*

The Heisenberg uncertainty principle imposes a limit to how accurately we can simultaneously measure the position and momentum of objects. The position of a particle is 'uncertain', because the act of observing it influences its position. The fundamental cause of this uncertainty is that photons transfer momentum to the object being measured as discussed in more detail below.

Since Heisenberg's uncertainty principle arises from photon momentum, it can be visualised using Nerf guns and an object of comparatively low mass such as a balloon. As shown in Figure 2, students hang balloons containing different masses of water. When students try to photograph a lowest mass balloon suspended by a string, the momentum of the bullets displaces the position of the balloon causing intrinsic fuzziness in the "image" of the balloon. On the other hand, when bullets hit the heaviest balloon, it hardly moves from its position. Similarly, when we try to measure the position and momentum of an object with light, the momentum from the light transfer to the object leads to uncertainty in both the position and momentum of the object. We have adapted this idea of momentum transfer into an engaging activity called the Nerf Gun Challenge (see Figure 2) that vividly illustrates the origin of the uncertainty principle.

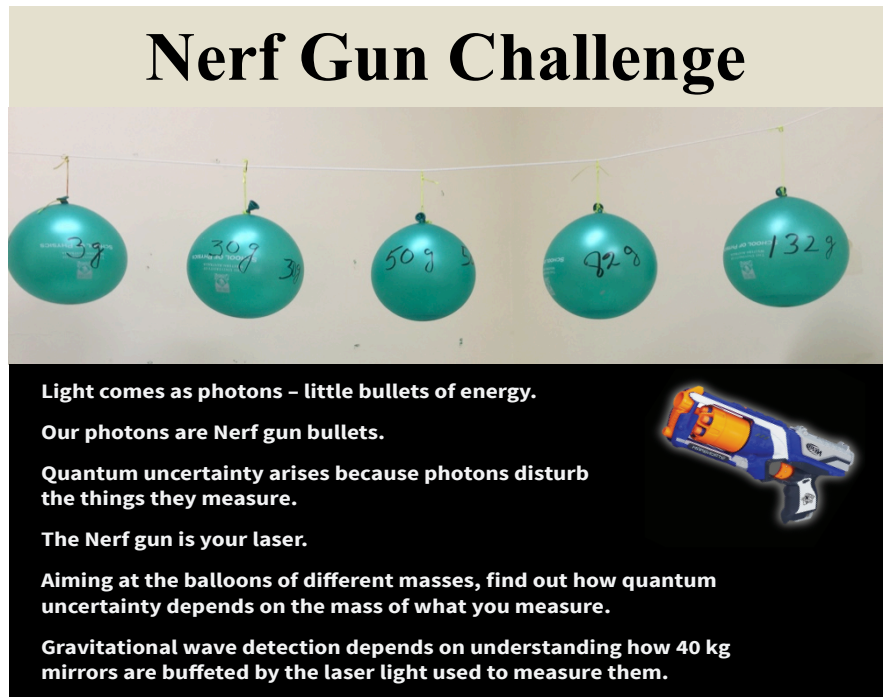

Figure 2. The ... physics involved. The balloons ... position oscillates as momentum is ... onal to the mass of the balloon; thus, the greatest oscillation is observed for the ... experiences negligible movement. This 'uncertainty' in the position of the balloon as a function of its size scales quite accurately to the magnitude of the uncertainty in the position of an atom when 'observed' by a photon.

The above argument is of course a simplification of the uncertainty principle. It can easily be extended (for more senior students) to include the standard quantum optical description of a laser interferometer gravitational wave detector.[14],[15] These devices achieve attometer ($10^{-18}$ m) measurement sensitivity to the displacement of 40 kg suspended mirrors, limited mainly by the uncertainty principle. The analysis of the origin of the uncertainty principle is suitable for presentation to students who have basic statistical knowledge about the sampling of random variables. The fundamental statistical concept is that when a random variable is sampled N-times, the standard error is $\sqrt{N}$. This corresponds to a 10% error for 100 samples, and 1% error for 10,000 samples etc. The fundamental quantum concept is that streams of photons are subject to the same statistical randomness as other random variables such as opinion poll surveys and coin tossing experiments.

Suppose an image such as the Nerf gun image of Figure 1 or the single photon interference video of reference 14(see section 2.6) is being used to make an optical measurement such as the mirror position in an interferometer. The mirror position changes the light intensity, so that the precision of the light intensity measurement determines the precision of the position measurement. The light intensity is determined by the number of photons arriving at a particular location in a specified time. However the photons arrive randomly, so if the mean number of photons is N, they will fluctuate by $\sqrt{N}$. This means that the fractional intensity fluctuation will be $\frac{\sqrt{N}}{N}$ or $\frac{1}{\sqrt{N}}$. Thus the estimate of the laser light intensity will improve the greater the number of photons. According to this, the measurement precision will improve indefinitely as the laser intensity is increased.

This is where photon momentum must be considered. Each photon exerts a kick to the mirror. If you make the intensity very high, N becomes very large. If the stream of photons had no randomness (i.e. it was a perfectly uniform stream of equally spaced photons) then you might imagine that the force could be rather steady. However as emphasised above, the photons arrive as a random stream, with intrinsic randomness in their arrival times. Thus for mean photon number N, the number of photons fluctuates by $\sqrt{N}$. Thus the momentum fluctuation acting on the mirror *increases* as $\sqrt{N}$. The random buffeting by the photons creates uncertainty in the mirror position (as observed in the Nerf gun experiment), and the more photons, the greater the uncertainty.

Combined together we have two effects that arise from the statistical nature of photon beams. The first, commonly called photon shot noise, *decreases* inversely as $\sqrt{N}$. The second, called radiation pressure noise, *increases* proportion to $\sqrt{N}$. The minimum uncertainty occurs when the radiation pressure uncertainty is equal to the shot noise uncertainty. This is commonly called the standard quantum limit, which is an expression of the uncertainty principle. Mathematically competent students can be asked to plot a graph of total uncertainty $A\sqrt{N} + \frac{B}{\sqrt{N}}$ (where A and B are arbitrary numbers chosen by the teacher) as a function of N, which represents light intensity. For all values of A and B there will be a characteristic minimum in uncertainty.

*2.5 The photoelectric effect with analogue photons*
The photoelectric effect occurs when electrons are ejected from the surface of a metal when irradiated by light of a certain frequency. The electrons of a metal are bound with a characteristic strength, which is known as its work function. The higher the work function of a metal, the stronger the bond between the metal and its electrons, and the more energy is required to sever this bond. Since the energy of light is proportional to its frequency, the light must be of a certain frequency to induce electron emissions. The minimum frequency of light required to eject an electron from a given metal is known as the threshold frequency. Only above this threshold does higher light intensity produce a higher current of photoelectrons. Technically if the quantum efficiency was unity, there would be one electron for every photon. We model this phenomenon in a simple interactive activity.

This activity will require a Nerf gun capable of firing a stream of bullets, bowls of varying depths and identical ping pong balls. In terms of our analogy, the Nerf gun bullets represent photons of a certain frequency (according to Planck's formula in Section 1), the bowls of ping pong balls represent the electrons bound inside the metal. The depth of the bowls represent the magnitude of the metal's work function (i.e. the strength of the bond between electron and metal).

To conduct the activity, arrange the bowls in a compact array and place a ping pong ball in the centre of each bowl as shown in Figure 3. Students stand a few meters from the array and prepare to fire their Nerf guns. Students can be asked to predict the consequences of firing Nerf gun bullets at the array- ping pong balls in shallower bowls are more likely to be ejected.

Since all Nerf gun bullets originate from the same source, we can assume they each have the same energy. In terms of our analogy, this corresponds to photons of the same frequency (and thus energy). Upon striking the ping pong ball, the bullet transfers some of its energy, causing the ball to oscillate in its bowl. In a shallow bowl (i.e. a metal with a low work function), this transfer of energy should be sufficient to eject the ball from the bowl (i.e. the threshold frequency of the metal has been exceeded). Ejected balls correspond to photoelectrons. For deeper bowls (i.e. metals with higher work functions), the threshold frequency will not be exceeded, and consequently the energy will not be sufficient to eject a photoelectron.

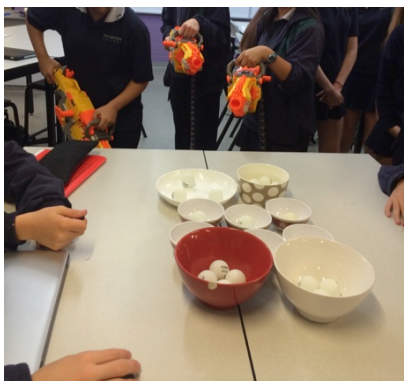

Figure 3. Students firing Nerf gun bullets at ping-pong balls resting in bowls of varying depths. Ping-pong balls in shallow bowls are ejected while those in larger bowls are not. This represents the photoelectric effect where electrons in metals with low work functions are ejected when irradiated by light exceeding the threshold frequency of the metal.

Using the same apparatus, we can assemble an alternative activity that emphasises the threshold frequency, rather than the work function. Arrange the bowls in a single of line of ascending depth, and place a single ping-pong ball in each. By firing nerf gun bullets at bowls in order of ascending depth, students can find the the bowl depth that corresponds to the threshold frequency for their Nerf gun photon energy.

*2.6 Single photon interference*
In Sections 2.7 – 2.8 we will present two classical wave activities with green laser pointers. But our purpose is not to prove that light is a wave but to uncover the hidden quantum reality. We do this by showing students videos of single photon interference experiments.[16] We ask students to note the random arrival of photons, and the final patterns which are compared with standard images from Young's double slit experiments.

Historically, before the discovery of the photon, interference experiments like those in the following sections were used to 'prove' that light is a wave. With the understanding that photons are spaced far apart (as we discussed in section 2.2), it is difficult to believe that the photons themselves are adding and subtracting like the ocean waves. According to student age, we use this opportunity to introduce the term wavefunction as the mathematical quantity that acts as if light was a continuous wave. It is the interference of this wavefunction that creates the pattern observed, because it's amplitude defines the probability that a photon will be detected at a certain point (see section 2.7 below for more). Yet this wavefunction does not appear to have physical existence: it is purely mathematical and is not physically observed. The following experiments are then presented in the context of the single photon interference.

*2.7 Diffraction of laser light through a human hair*
Laser diffraction and interference are normally used to illustrate wave-like behavior of light. However, since we have already emphasised that light comes as photons, as well as the concept of the spacing between photons, these activities also reveal the weirdness of quantum mechanics, as discussed in this section, and in more depth in section 2.8.

Prior to commencing this activity, we first illustrate classical diffraction by showing students photographs of ocean waves diffracting around islands (see Figure 4(a) below). We explain that waves from either side of the islands, redirected by the landmass, interfere with each other and create inference patterns as the waves add and subtract.

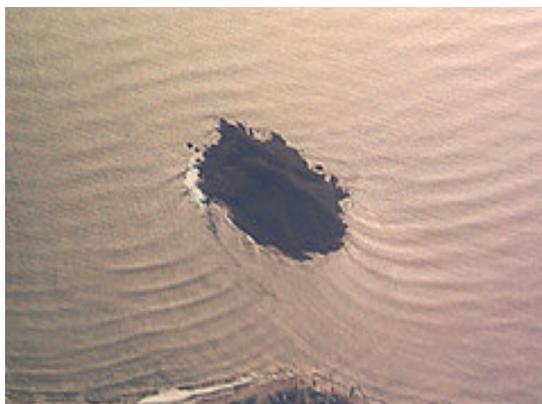

Figure 4(a).[17] Water waves diffracting around a small island, creating interference when the waves combine on the other side of the island. Patterns of large and small amplitude are observed as the waves approach the shoreline.

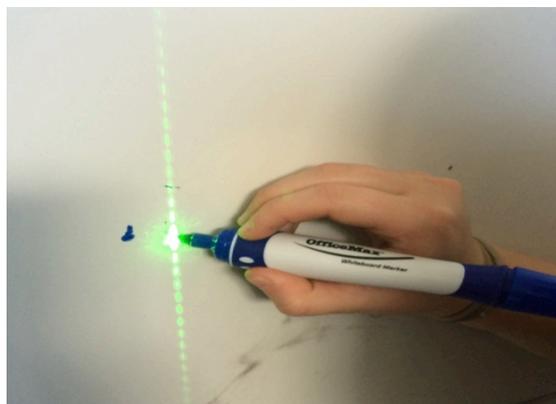

Figure 4(b). Vertical interference pattern created by the diffraction of green light through a horizontal human hair. The bright and dark bands of light represent regions of constructive and destructive respectively. These regions can be marked and measured to calculate the diameter of the hair around which the light has diffracted.

Using this illustration as an analogy for the upcoming activity, we substitute a laser beam (a stream of equal energy photons) as the wave, a single human hair as the island, and a screen to display the wave intensity.

A single human hair is placed in the beam of a green laser pointer, and directed to a screen 5-10 meters away. Students observe an intensity pattern as illustrated in Figure 4(b). By comparing the interference pattern created by the laser to that in the image of ocean waves, students can appreciate that the photons create a brightness distribution similar to that of a wave.

Students are then asked to calculate the width of their own hair by measuring the spacing between successive bands in the interference pattern (see Figure 4(b)), and employing the following simple formula

$$d = \frac{\lambda L}{x}$$

where, d is the hair diameter, $\lambda$ is the wavelength of the laser light, L is the distance from hair to the whiteboard, and x is the distance between successive light/dark bands.[18] For junior high school students we simplify the formula to hair diameter (microns) = C/fringe spacing (cm), where the constant C is evaluated for the laser wavelength (preferably 532nm) and screen distance.

Just as the interference pattern created by the island depends on both the size of the island and the wavelength of the waves, the laser interference pattern depends on both the diameter of the hair and the wavelength of the laser light. Students can compare the thickness of their hair with their classmates'.

*2.8 Soap film interference*
Another vivid and beautiful illustration of the wave-like behaviour of quantum interference can be seen using reflective interference of light in soap bubble films. Light reflected from the front and rear surfaces of a soap film produces high contrast interference patterns that can be projected onto a screen (see Figure 5). This interference pattern arises from the phase difference between light reflected from either side of the soap film wall.

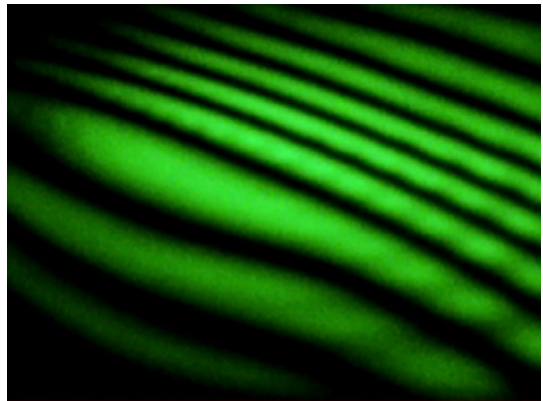

Figure 5. Interference pattern created by green light reflected from a soap bubble. Light reflected from either side of the soap film wall interferes, creating high contrast characteristic regions of constructive and destructive interference. Despite the solidity and permanence of the interference pattern, for moderate light intensity, it is easy to show that the image is created by single isolated photons in the soap bubble film.

The concept is illustrated in the following two diagrams (see Figures 6(a) and 6(b) below): the first shows what you would expect if light were a continuous wave. The second shows how we may imagine it actually occuring in the case of discrete photons. The fact that interference occurs, even with a very dim light source, when the spacing between photons greatly exceeds the thickness of the soap film, demonstrates the weirdness of the quantum world.

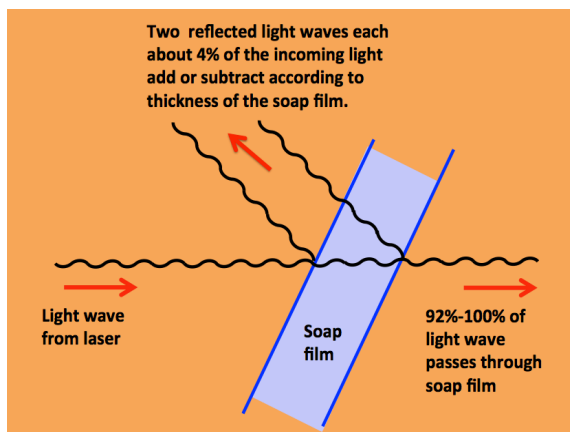

False classical description of soap film interference

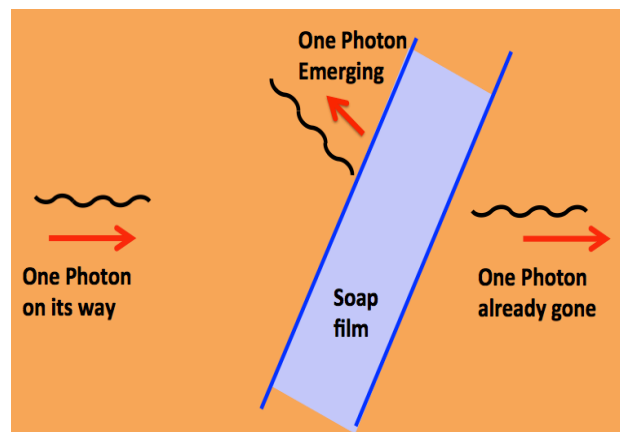

Quantum description of soap film interference

Figure 6(a). Light, treated as a continuous wave, reflected from a soap film. While the majority (92-100%) of the wave will pass through the film unperturbed, a small fraction is reflected from each surface of the film. The reflected waves with reduced amplitude will interfere, producing characteristic patterns of constructive and destructive interference as shown in Figure 5.Historically such patterns were used to prove that light is a wave.

Figure 6(b). Light, treated as photons (discrete packets of, energy) reflected from a soap film. Unlike a continuous classical wave that can have arbitrary amplitude, photons, being discrete packets of energy, cannot have reduced amplitude. Yet characteristic wave interference patterns still emerge. The weirdness of quantum behavior is evident when one realizes that at modest light intensity only one photon is present in the soap film at any time.

The source of the interference is the soap film wall, a structure of only a few micrometers thickness. Since we have already calculated in Section 2.2 that individual photons are separated by distances often measured in kilometers, we are presented with a puzzling scenario. At any instant, there can never be more that one photon present inside the soap film, yet the interference pattern that emerges appears permanent. If the interference pattern is not caused by photons, what is the cause? As discussed in Section 2.6 the interference is explained as interference of a mathematical entity called the wavefunction, which determines the probability that a photon will arrive at a particular location. The probability follows exactly the same mathematics that the intensity would follow for a real continuous wave. It is the wavefunctions that interfere, not the photons themselves. Dark parts of the interference pattern are places where the wavefunction interference causes the probability of a photon arriving to be zero. The bright parts are places where the wavefunction interference is maximum. This single-particle interference is one of the hallmark phenomena of quantum physics and a strong demonstration of 'quantum weirdness'. Again we emphasise that wavefunctions are mathematical entities that cannot be measured or seen except for their effects on the probability of photons arriving at a particular location.

*2.9 Limitations of this quantum model*
We have used Nerf gun bullets as models for photons. While this model is useful for emphasising the particle nature of light, its limitations must be emphasised. These limitations are in the sense that the correspondance between model and reality breaks down. Identifying precisely where and how the analogy breaks down, in a thorough discussion with students, is important for obtaining a deep understanding of the concepts discussed above. Students deepen their understanding by discussing the failing of the model.

The representation of photons by toy bullets is very useful for illustrating the discreteness, the energy and the momentum of photons, and phenomenon such as the photoelectric effect and the uncertainty principle. However obviously, Nerf gun bullets have non-zero rest mass and cannot propagate at the speed of light, nor can they produce interference patterns. Their energy is dependent on their velocity, and their trajectories are strongly distorted by gravity. These obvious limitations can be identified by the students in a group discussion.

The discussion about the spacing between photons assumes that photons are really particle-like. This is not necessarily true for real photons. The reality is that physicists do not have a clear, simple understanding of what a photon is like. You cannot image a photon. When you detect a photon it is detected like a particle, but before you detect it we cannot say what it is like. Students need to know that quantum mechanics is a mathematical theory. It has been tested to enormous precision and proved to be correct. However to this day, physicists argue about the meaning of quantum reality. Richard Feynman said that if you meet a person who says they understand quantum physics, they are either stupid or lying! Yet because the theory works perfectly, we have been able to use it to design the components of computers, lasers and mobile phones, and without the theory we would never have been able to create such marvellous devices.

**3. Conclusion**
The quantum physics component of the Einstein-First project combines analogies with thought experiments to reveal the weirdness of quantum behaviour. We have presented an observation based approach which allows some of the key concepts of quantum physics to be introduced at an early age. It allows students to understand quantum uncertainty, and to realize the intrinsic weirdness of quantum reality as a fact of nature that we must accept.

In modern text books and school curricula the term wave-particle duality is used, which tends to imply equal value for the wave and the particle properties. We have emphasised that modern experiments always observe light as a particle. Photons in starlight may be 1000km apart, and single photon experiments show images of interference growing one photon at a time. This implies that the wave-nature of light, as supposedly revealed by interference experiments is to some extent illusory.

We presented a class activity that drew a parallel between a classical water wave phenomenon and laser interference experiment, but emphasised that the pattern is unchanged when photons arrive one at a time. We showed that this pattern informs us about the apparatus (a human hair) and the photon wavelength (or frequency or energy). Thus, while the particle nature of photons is inescapable, the wavelike aspect of photons is also inescapable.

The weirdness of quantum mechanics arises because we do not understand why the arrival probability of quantum particles is described by hypothetical wavefunctions that have never been observed, but which are defined by directly measurable wave-quantities such as wavelength, as if it was a classical wave. Part 3 of this series reviews some of the research results obtained from all our programs. In particular it gives results of testing which was used to assess student understanding of the quantum physics concepts discussed here.

**4. Acknowledgements**
This research was supported by the Australian Research Council, the Gravity Discovery Centre and the Graham Polly Farmer Foundation.

**Appendix**
Here we outline the materials required to construct the various models and activities described in this paper:

**1) Photography with analogue photons:** A whiteboard, whiteboard markers to outline the images, Nerf guns (we used the model vulcan EBF-25), Nerf gun bullets with suction ends, and safety glasses.

**2) Uncertainty principle with analogue photons :** Balloons filled with 2g -150g of water**,** a Nerf gun, Nerf gun bullets and string to hang the balloons.

**3) Photoelectric effect with analogue photons:** Approximately 10 bowls with 6-20cm depths to represent different work functions. A few ping pong balls in each bowl represent the electrons in the metal. Several Nerf guns with foam bullets are sufficient for a class activity..

**4) Laser diffraction of a human hair:** 1 mW green laser module or a laser pointer, a ruler, a marker, a white screen and a human hair.  A cardboard box with a cut out and double sided adhesive tape to suspend hair samples across the hole. Care must be taken to avoid directly shining laser light at students' faces.

**5) Soap bubble interference:** 1 mW green laser module, a screen and a soap bubble frame consisting of a malleable copper wire moulded in the shape of a 1-2cm diameter ring that can be dipped into a soap solution. The wire ring should be twisted out of the plane so as to create a convex soap film surface. This causes magnification of the reflected interference pattern.

# References

[1] Hertz, H. (1887). "Ueber den Einfluss des ultravioletten Lichtes auf die electrische Entladung" [On an effect of ultra-violet light upon the electrical discharge]. *Annalen der Physik*. **Vol. 267 (8)**, p.S983–S1000.

[2] Einstein, Albert (1905). "Über einen die Erzeugung und Verwandlung des Lichtes betreffenden heuristischen Gesichtspunkt". *Annalen der Physik*. 17 (6): 132–148. http://myweb.rz.uni-augsburg.de/~eckern/adp/history/einstein-papers/1905_17_132-148.pdf Retrieved 28-08-2016.

[3] Feynman, R. (1985). QED: The Strange Theory of Light and Matter (Alix G. Mautner Memorial Lectures) (New Jersey: Princeton University Press. p.15.

[4] Mori, O. et al. (2009). "First Solar Power Sail Demonstration by IKAROS" (PDF). *27th International Symposium on Space Technology and Science.* Retrieved 13 June 2017.

[5] Henriksen, E. et al. (2014). "Relativity, quantum physics and philosophy in the upper secondary curriculum: challenges, opportunities and proposed approaches", *Physics Education*, **Vol. 49**, p.678.

[6] Johansson, K. & Milstead, D. (2008). "Uncertainty in the classroom—teaching quantum physics", *Physics Education*, **Vol. 43**, p. 173.

[7] Hadzidaki, P., Kalkanis, G., and Stavrou, D. (2000). "Quantum mechanics: a systemic component of the modern physics paradigm", *Physics Education,* **Vol. 35**, p.386.

[8] Gordon, M. and Gordon, G. (2012). "Quantum computer games: Schrödinger cat an hounds", *Physics Education,* **Vol. 47**, p. 346.

[9] Malgieri, M., Onorato, P., and Ambrosis, A. (2014). "Teaching quantum physics by the sum over paths approach and GeoGebra simulations", *European Journal of Physics*, **Vol. 35**.

[10] Singh, C. (2008). "Interactive learning tutorials on quantum mechanics", *American Journal of Physics*, **Vol. 76**.

[11] Müller, R. and Wiesner, H. (2002). "Teaching quantum mechanics on an introductory level", *American Journal of Physics*, **Vol. 70**, p. 200.

[12] Tinsley, J. N. *et al.* (2016). *Nature Communication.* **7**, 12172

[13] https://www.youtube.com/watch?v=MbLzh1Y9POQ

[14] Blair et al. (2012). Advanced gravitational wave detectors, p.63.

[15] Jaekel, M. and S. Reynaud, S. (1990). "Quantum Limits in Interferometric Measurements", Europhysics Letters, **Vol. 12(4)**.

[16] http://photonterrace.net/en/photon/duality/

[17] https://www.flickr.com/photos/physicsclassroom/galleries/72157625274620927/

[18] http://physics4chynyein.blogspot.com.au/2012/04/experiment-11-measuring-human-hair.html